\documentclass[10pt,a4paper]{article} % 10pt is ignored!
\usepackage[dvips]{graphicx}
\usepackage{amsfonts,amsmath,amssymb}
\usepackage{hyperref}
\usepackage{cite}
\newcommand{\bse}{\begin{subequations}}
\newcommand{\ese}{\end{subequations}}

\begin{document}
\title{\bf Holographic Schwinger Effect in a D-Instanton Background}

\date{}
\maketitle
\vspace*{-0.3cm}
\begin{center}
{\bf Leila Shahkarami$^{a,1}$, Majid Dehghani$^{b,2}$, Parvin Dehghani$^{c,3}$}\\
\vspace*{0.3cm}
{\it {${}^a$School of Physics, Damghan University, Damghan, 41167-36716, Iran}}\\
{\it {${}^b$Research Institute for Astronomy and Astrophysics of Maragha (RIAAM), P. O. Box 55134-441, Maragha, Iran
}} \\
{\it {${}^c$School of Mathematics and Statistics, 4302 Herzberg Laboratories, Carleton University, 1125 Colonel by Drive, Ottawa, Ontario, K1S 5B6, Canada
}}  \\
\vspace*{0.3cm}
{\it  {${}^1$l.shahkarami@du.ac.ir}, {${}^2$m.dehghani55@gmail.com}, {${}^3$parvindehghani@gmail.com} }
\end{center}

\begin{abstract}
The Schwinger effect in the presence of instantons is considered in this paper. Using AdS/CFT correspondence in the near horizon
limit of the D3+D($-1$)-brane background, we calculate the total potential of a quark-antiquark pair in an external electric field.
It is shown that instantons tend to suppress the pair creation effect and increase the critical electric field above which the pairs are produced freely without any suppression. 
Interestingly, no other critical electric field, common for all confining field theories, is observed here at finite temperature.
However, as expected we find such a critical electric field at zero temperature.
The pair production rate evaluated by the calculation of the expectation value of the circular Wilson loop also confirms this result.
\end{abstract}
Keywords: Schwinger effect; AdS/CFT; Confinement; Entanglement entropy.
%$$$$$$$$$$$$$$$$$$$$$$$$$$$$$$$$$$$$$$$$$$$$$$$$$$$$$$$$$$$$$$$$$$$$$$
\section{Introduction}
The electron and positron pair production in the presence of an external electric field in the vacuum of the quantum electrodynamics (QED) is a nonperturbative phenomenon known as the Schwinger effect \cite{Schwinger}.
This effect is not restricted to QED, but it is also relevant to QFTs coupled to a $U(1)$ gauge field.
The original Schwinger's work was based on the weak-coupling and weak-field approximation.
He found an exponential suppression with a quantity $\propto \frac{m^2}{e E}$ in the exponent for the production rate, in the presence of the external field $E$, where $m$ and $e$ denote the rest mass and electric charge of the electron. 
Until now, the Schwinger effect has not been observed in real experiments, since a strong electric field, greater than $\frac{m^2}{e}\approx 1.3\times 10^{12} V/m$, is needed for this effect to become significant and observable.

Later, this effect was generalized to the arbitrary-coupling but weak-field case \cite{manton1}, by taking into account the Coulomb interaction.
Modifying the potential in this way leads to the existence of a critical electric field $E_c$ below which the pair production can be explained as a tunneling process; that is, the particles are faced with a potential barrier.
As the electric field increases to $E_c$, the potential barrier vanishes and above $E_c$ the production rate is not suppressed anymore and hence the vacuum becomes completely unstable.
However, in QED the value obtained for the critical electric field does not satisfy the weak-field condition.
Therefore, to verify the existence of the critical value, we need to work beyond this condition.

The existence of a critical value for the electric field and consequently a phase transition is predicted by the string theory \cite{string1,string2}.
The connection between the string theory and gauge theories is established through AdS/CFT correspondence \cite{Maldacena1,Maldacena2,Maldacena3,Solana}.
This connection motivated people to study the Schwinger effect in the context of AdS/CFT and more generally gauge/gravity duality, in which the problem can be considered beyond the weak-field approximation.

The ${\cal N}=4$ super-Yang-Mills (the theory in the field theory side of AdS/CFT) does not contain matter fields.
One way to introduce them to the theory is to employ the Higgs mechanism in which the gauge group is broken from $SU(N+1)$ to $SU(N)\times U(1)$. 
Using the scheme of \cite{manton1} in the large $N$ limit, the production rate for this theory can be obtained.
However, this approach leads to a puzzle. 
The critical electric field obtained in this way disagrees with the one derived by the Dirac-Born-Infeld (DBI) action of a D3-brane near the anti-de Sitter (AdS) boundary.
This problem was resolved by the holographic setup proposed by Semenoff and Zarembo \cite{semenoff}.
Considering a single probe D3-brane in an intermediate position in the bulk and separated from the stuck of $N$ D3-branes, they built a setup in which the mass of a single quark is finite.
Then, they evaluated the production rate by computing the expectation value of a circular Wilson loop on the probe brane, using the holography dictionary, and found a value for the critical electric field in agreement with the DBI action result.
Then, a consistent potential analysis was invented \cite{potential} by using a modified Coulomb potential which gives results in agreement with the previous ones.
The potential between a particle-antiparticle pair at rest comes from the expectation value of the rectangular Wilson loop which corresponds to the area of a string world sheet attached to the Wilson loop on the boundary.

Since then, there has been a growing interest in investigating various aspects of the holographic Schwinger effect in different situations (for example see \cite{Sch1,Sch2,Sch3,confin1,confin2,Sch4,Sch5,Sch6,Sch7}).
Here we are interested in studying the Schwinger effect in confining gauge theories using holography, originated from \cite{confin1}.
A review on this topic can be found in \cite{confinrev}.
The study of this effect in confining backgrounds may shed some light on the confinement/deconfinement phase transition in QCD.
Especially, this effect is relevant to the heavy ion collisions, done in, e.g., the RHIC and LHC experiments, where strong electric and magnetic fields are present.

In the present paper we consider a quasiconfining gauge theory at both zero and finite temperatures, and explore the effect of instantons on the Schwinger effect in this background.
To do so, we calculate both the potential and the pair production rate, and discuss about  the possible phases of the system.
It is known that Yang-Mills instantons are identified with the D-instantons of type IIB string theory (for example see \cite{dual-ins1,dual-ins2,dual-ins3}).
We choose the D3+D($-1$)-brane background for our purposes.
This background at zero temperature was first suggested in \cite{instanton1}.
They discussed the near horizon limit of the D-instanton charge homogeneously distributed over the D3-brane world volume at zero temperature
and proposed that it corresponds to a $\mathcal{N}=4$ super-Yang-Mills theory in a constant homogenous self-dual gauge field background, where the presence of a self-dual gauge field in flat space is dual to the presence of the D-instanton charge. 
Using the AdS/CFT correspondence, they also showed that this background should be a partially confining theory with confined quarks and deconfined gluons and this is the reason why we use the term quasiconfining for this theory. 
%This corresponds to nonzero vacuum expectation value of gluon condensation $q \propto\langle Tr F^{\mu\nu}F_{\mu\nu} \rangle\neq0 $, responsible for chiral symmetry breaking. 
Also, the holographic dual of a uniformly distributed D-instanton over the D3-brane at finite temperature for D7-brane embedding was considered in \cite{instanton2}.

Although the holographic Schwinger effect in confining gauge theories has been studied in different backgrounds and universal behaviors have been obtained \cite{Sch1}, the theory chosen here is different in that it has confined quarks but deconfined gluons at zero temperature and may help us understand new things about the confinement.
Also, this theory does not have any compactified direction and therefore shows no geometric transition as temperature grows, unlike the other confined theories studied under an external electric field.
%  The above mentioned features of D3-D(-1) brane configuration is similar to QCD at finite temperature. From lattice QCD calculations and  heavy ion collision experiment it is known that heavy quark mesons (quarkonium) survive above the transition temperature to quark-gluon plasma (QGP). So the finite temperature QCD and D3-D(-1) configuration are partially confined theories for heavy quarks.

In the next section the background geometry of D3+D($-1$)-brane configuration at finite temperature is introduced.  
Considering a string hanging from the boundary of the above mentioned background, the total potential, including the static energy, the Coulomb potential energy, and the potential energy due to the interaction with the external electric field is computed in Sec. 3.
Moreover, using the DBI action of the probe D3-brane, a critical field for pair production is obtained, and used to confirm the correctness of the potential method used here.
To show the effect of instantons on the process, we do the calculations for different values of the instanton density parameter and for both zero and finite temperatures. 
We devote Sec. 4 to investigate the confinement of the system using the notion of the pair production rate.
We finally summarize and draw our conclusions in Sec. 5.

%$$$$$$$$$$$$$$$$$$$$$$$$$$$$$$$$$$$$$$$$$$$$$$$$$$$$$$$$$$$$$$$$$$$$$$$$$
\section{Background geometry with D-instantons}\label{backgr geo}
This section is devoted to a short introduction of the background with D-instantons in the gravity side. 
We are interested in the near horizon limit of D3+D($-1$)-brane geometry at finite temperature with the Euclidean signature.
The ten-dimensional supergravity action in Einstein frame is given by \cite{instanton2}
\begin{align}\label{action10D}
 S=\frac{1}{\kappa}\int d^{10}x \sqrt{g} \left(R-\frac{1}{2}(\partial\Phi)^2 +\frac{1}{2}e^{2\Phi}(\partial\chi)^2-\frac{1}{6}F^2_{(5)} \right),
\end{align}
where $\Phi$ and $\chi$ are the dilaton and axion fields, respectively, and $F_{(5)}$ denotes a five-form field strength. 
If we set $\chi=-e^{-\Phi}+\chi_0$, the dilaton and axion terms cancel each other in the above action. 
Then, the solution with the metric and five-form field in the string frame is obtained in the following form:
\begin{align}\label{action22}
 ds^2_{10}&=e^{\Phi/2} \left\{\frac{r^2}{R^2}\left[f(r)^2 dt^2+d\vec{x}^2\right]+\frac{1}{f(r)^2}\frac{R^2}{r^2}dr^2+R^2d\Omega^2_5 \right\}, \nonumber  \\
 e^\Phi&=1+\frac{q}{r^4_T}\log\frac{1}{f(r)^2},~~~~ \chi=-e^{-\Phi}+\chi_0, \nonumber  \\
 f(r)&=\sqrt{1-\left(\frac{r_T}{r}\right)^4}.
 \end{align}
Here, $R$ and $r_T$ are the radius of the AdS space and the event horizon, respectively.
$r_T$ is related to the temperature of the dual gauge theory.
Moreover, $q$ denotes the density of D-instantons, which corresponds to the vacuum expectation value of the gluon condensation according to the AdS/CFT dictionary. 

In the following section we consider the Schwinger effect in this background and especially we focus on the effects coming from $q$ related to the presence of the D-instantons.
%*********************************************************************
\section{Potential analysis}
In this section, we want to perform the potential analysis in the aforementioned background.
To do so, we should calculate the total potential for a quark-antiquark pair, including the potential energy ($V_{\mathrm{PE}}$), the static energy ($V_{\mathrm{SE}}$), and the energy of interaction with the external electric field $E$.
However, before that we need to find the critical value of the electric field obtained from the DBI action of the probe D3-brane. 
This calculation helps to ensure that our potential analysis is correct and agrees with the DBI result.

The DBI action of a probe D3-brane in the D3+D($-1$) background, located at $r=r_0$ and including a constant world-volume electric field, is of the form
\begin{align}\label{DBI action}
 S_{\mathrm{DBI}}&=-T_{D3}\int d ^4x \sqrt{-\det(g_{\mu\nu}+\mathcal{F}_{\mu\nu})}     \nonumber \\
        &=-T_{D3} \frac{r_0^4}{R^4}e^{\Phi(r_0)/2}\sqrt{1-\frac{r^4_T}{r^4_0}} \int d^4x \sqrt{1-\frac{(2\pi \alpha')^2 R^4}{r^4_0 e^{\Phi(r_0)}(1-r^4_T/r^4_0)}E^2},
\end{align}
where $T_{D3}=1/({g_s(2\pi)^3 \alpha'^{2} })$ is the D3-brane tension.
It can easily be seen from this equation that the classical solution does not exist for $E>E_c$, where the critical electric field $E_c$ is 
\begin{equation}\label{critical field}
  E_c=\frac{1}{2\pi\alpha'}\frac{r_0^2}{R^2}\sqrt{e^{\Phi(r_0)} \left(1-\frac{r^4_T}{r^4_0} \right)},
\end{equation}
which obviously depends on the temperature through $r_T$ and on the instanton density through $\Phi(r_0)$.

Now, we proceed to the calculation of the total energy.
To that purpose, we consider a quark and an antiquark placed at fixed positions on the boundary, separated by a distance $x$ in one of the three spatial directions of the field theory, e.g., $x_1$.
In order to use the symmetry, we choose $x_1=0$ to be halfway between the quark and antiquark.
%The quark-antiquark pair corresponds to a string connecting their positions in the gravity side. 

In order to obtain the potential energy between the quark-antiquark pair, one needs to compute the expectation value of the rectangular Wilson loop. 
It is well known that when the time duration, $\tau$, is much larger than the separation $x$, the Wilson loop expectation value takes the form 
\begin{equation}\label{static gauge}
  \langle W \rangle=e^{-i (2 m +V(x))\tau},
\end{equation}
where $m$ denotes the quark and antiquark rest mass and $V(x)$ is the potential energy between them.
In fact the exponent is the static energy plus the potential energy, $V_{\mathrm{PE}+\mathrm{SE}}$.
In order to have finite-mass quarks, following \cite{semenoff}, we put a probe D3-brane at an intermediate position $r_0$ in the bulk and attach the endpoints of the string to this brane.
In the holographic setup the exponent in Eq.\,(\ref{static gauge}) corresponds to the world-sheet area or equivalently the on-shell Nambu-Goto (NG) action of the hanged string.
The use of the parametrization of the string coordinates as $x^0=\tau$ and $x^1=\sigma$ (the static gauge), and supposing $r=r(\sigma)$, give rise to the following form for the string action:
\begin{equation}\label{action}
  S_{\mathrm{NG}}=T_f\int d\tau d\sigma\sqrt{e^{\Phi(r)}\left[\left(\frac{dr}{d\sigma}\right)^2+\frac{r^4}{R^4}\left(1-\frac{r^4_T}{r^4}\right)\right]},
\end{equation}
where $T_f=\frac{1}{2 \pi \alpha'}$ is the string tension.
The independence of this action from the coordinate $\sigma$ introduces a conserved quantity (Hamiltonian) which is written as follows:
\begin{equation}\label{Hamiltonian}
  \frac{e^{\Phi(r)} r^4/R^4}{\sqrt{e^{\Phi(r)} \left[\left(\frac{dr}{d\sigma}\right)^2+\frac{r^4}{R^4}\left(1-\frac{r^4_T}{r^4}\right) \right]}} \left(1-\frac{r^4_T}{r^4} \right)=\mathrm{const}.
\end{equation}
One can determine the constant in this relation using the conditions at $\sigma=0$, i.e.,\;$r'(0)=0$ and $r(0)=r_c$, which results in the following differential equation:
\begin{equation}\label{drdsigma}
  \frac{dr}{d\sigma}=\frac{r^2}{R^2} \sqrt{\left(1-\frac{r^4_T}{r^4}\right) \left(\frac{e^{\Phi(r)}(r^4-r^4_T)}{e^{\Phi(r_c)}(r^4_c-r^4_T)}-1 \right)}.
\end{equation}
Integrating this equation, we obtain the separation length of the quark-antiquark pair on the probe brane as follows:
\begin{align}\label{lenthx}
  x =& \frac{2 R^2}{r_0 a} \sqrt{e^{\Phi(1)}(1-b^4/a^4)} \nonumber \\
   &\times  \int^{1/a}_1 \frac{dy} {\sqrt{ \left(y^4-b^4/a^4\right)\left[e^{\Phi(y)}\left(y^4-b^4/a^4\right)-e^{\Phi(1)}\left(1-b^4/a^4\right) \right]}},
\end{align}
in which the following dimensionless quantities have been introduced,
\begin{equation}\label{parameters}
  y\equiv\frac{r}{r_c}, ~~~~ a\equiv\frac{r_c}{r_0},  ~~~~ b\equiv\frac{r_T}{r_0},
\end{equation}
and
\begin{equation}\label{ephi change}
 e^{\Phi(y)}=1+\frac{q}{r^4_0 b^4}\log \left(\frac{y^4}{y^4-b^4/a^4}\right).
\end{equation}
Then, by substituting Eq.\,(\ref{drdsigma}) into Eq.\,(\ref{action}), the sum of potential and static energy, $V_{\mathrm{PE}+\mathrm{SE}}$, can be obtained.
Then, we obtain the total energy by adding the potential energy due to the interaction of the quark-antiquark pair with an external electric field $E$ to this energy, which can be written as follows:
%\begin{equation}\label{CPandSE}
 % V_{CP+SE}=\int^{1/a}_1 dy e^{\Phi(y)}r_c \sqrt{ \left( y^4-\frac{b^4}{a^4} \right) \left(\frac{y^4-1}{e^{\Phi(r_c)}(1-b^4/a^4)}+1 \right)}.
%\end{equation}
%Now, the total potential can be obtained by adding the above energy to the potential energy due to the interaction of the quark-antiquark pair with an external electric field $E$, which can be written as follows
\begin{align}\label{total poten}
  V_{\mathrm{tot}}(x)&=V_{\mathrm{PE}+\mathrm{SE}}-Ex         \nonumber \\
        &=2 \ T_f\int^{1/a}_1 dy  \frac{r_0 a e^{\Phi(y)}(y^4-b^4/a^4)-\frac{\alpha}{a}r_0 \sqrt{e^{\Phi(1)} e^{\Phi(1/a)} (1-b^4)(1-b^4/a^4)}}{ \sqrt{\left(y^4-b^4/a^4\right)\left[e^{\Phi(y)}\left(y^4-b^4/a^4\right)-e^{\Phi(1)}\left(1-b^4/a^4\right) \right]}},
 \end{align}
where the dimensionless parameter $\alpha$ is defined as $ \alpha=\frac{E}{E_c(T,q)}$ using the critical electric field obtained from the DBI result [Eq.\,(\ref{critical field})].

Now, we report the results.
In all the following cases we set $T_f r_0=1$.
%==================================================
\subsection{Potential analysis at finite temperature}
\begin{figure}[h]
\begin{center}
\includegraphics[width=5cm]{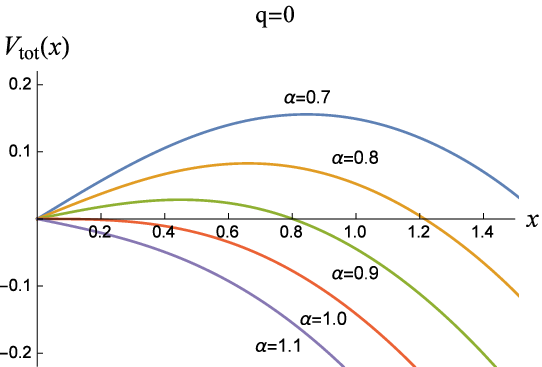}\hspace{.1cm}
\includegraphics[width=5cm]{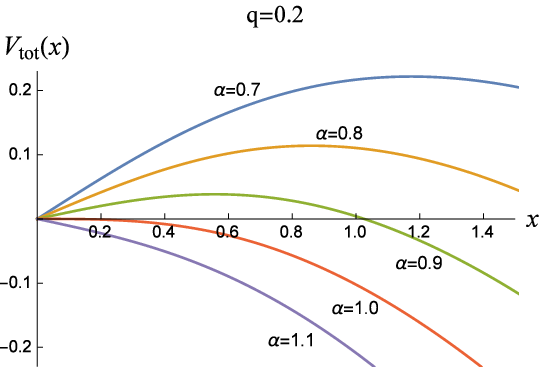}\hspace{.1cm}
\includegraphics[width=5cm]{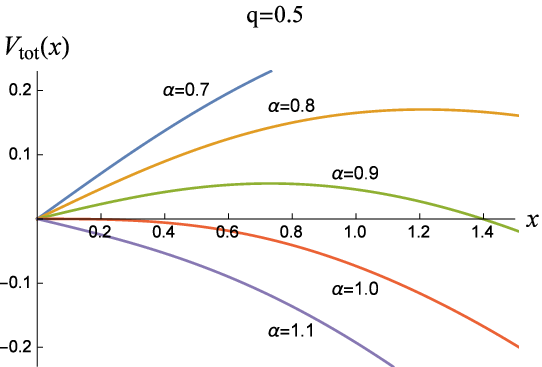}
\end{center}
\caption{\footnotesize 
The total potential versus $x$ for various instanton densities and electric fields (determined by $\alpha$).
In all graphs $b=0.5$.}
\label{alpha-V}
\end{figure}

\begin{figure}[h]
\begin{center}
\includegraphics[width=6cm]{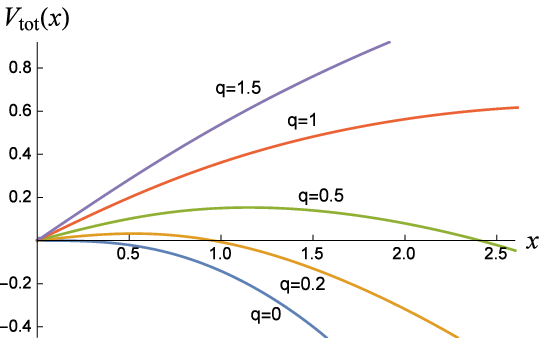}
\end{center}
\caption{\footnotesize 
The total potential versus $x$ at the critical electric field and for various instanton densities.
In all graphs $b=0.5$ and $\tilde{\alpha}=1$.}
\label{v-tilde}
\end{figure}

Figure \ref{alpha-V} shows the total potential versus the separation $x$ for a fixed value of $b$ (fixed temperature) and different values of $\alpha$ (namely, $\alpha$=0.7, 0.8, 0.9, 1, 1.1).
The instanton density has been chosen to be different in the three panels of this figure, for comparison. 
These graphs show that the potential barrier vanishes for $\alpha > 1$, confirming that the critical electric field obtained from the potential analysis agrees with the one of the DBI result, as expected \cite{Sch1}.
By comparing the results we can conclude that instantons increase the potential barrier and suppress the pair creation.
However, the critical field ($E_c$) and consequently the parameter $\alpha$ depend on the instanton density $q$.
Therefore, in order to explore the instanton effect more clearly, we introduce another dimensionless parameter as $\tilde{\alpha}=\frac{E}{E_c(T,q=0)}$.
In Fig.\,\ref{v-tilde} the electric field is chosen to be equal to the critical electric field for zero instanton density, $E_c(T,q=0)$ which corresponds to $\tilde{\alpha}=1$. This figure shows that at zero instanton density there is no potential barrier in the presence of this electric field; that is, the pair creation happens without any limitation.
However, the presence of the instantons develops a potential barrier.
Therefore for nonzero $q$ the Schwinger effect occurs only through a tunneling process and a larger instanton density leads to a larger potential barrier. 
This means that the critical electric field is increased by rising $q$ from zero.
Notice that this result is obvious from the relation (\ref{critical field}).
 
%==================================================
\subsection{Potential analysis at zero temperature}
\begin{figure}[h]
\begin{center}
\includegraphics[width=6cm]{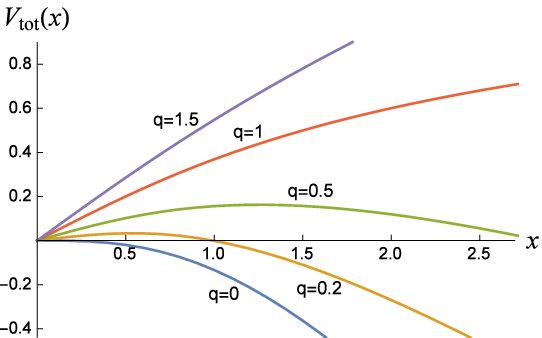}
\end{center}
\caption{\footnotesize 
The total potential versus $x$ at the critical electric field and for various instanton densities.
In all graphs $b=0$ and $\tilde{\alpha}=1$.}
\label{ft0}
\end{figure}

All the previously found results for a general $b$ can be reduced to the zero temperature case by setting $b=0$.
Notice that in this case $e^{\Phi(y)}=1+\frac{q}{r^4_c y^4}$.
The total potential as a function of $x$ is shown  in Fig.\,\ref{ft0} for various values of the instanton density at zero temperature.
In this figure the electric field is equal to its critical value with $q=0$.
Again we see that increasing the instanton density increases the potential barrier as in the case of finite temperature and therefore the value of $E_c$ increases by increasing $q$.

\begin{figure}[h]
\begin{center}
\includegraphics[width=6cm]{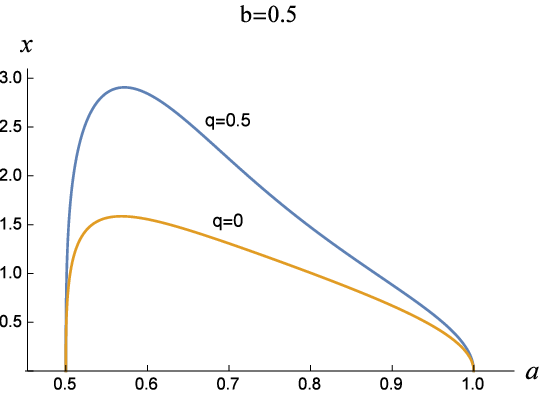}\hspace{.5cm}
\includegraphics[width=6cm]{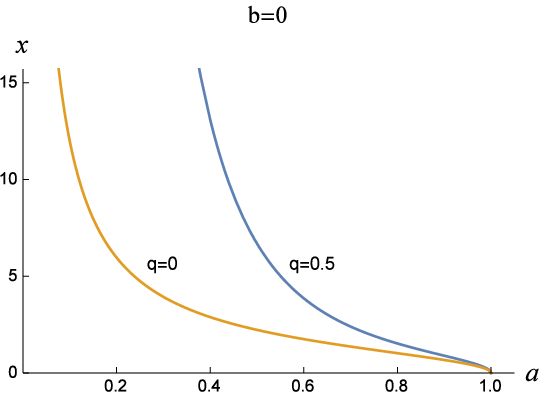}
\end{center}
\caption{\footnotesize 
The separation length of the quark-antiquark pair versus the rescaled position of the turning point of the corresponding string in the bulk.}
\label{x-a}
\end{figure}

An interesting result is that in the finite temperature case the theory behaves like a deconfined theory as viewed by an external electric field, since there is no critical electric field below which the Schwinger effect is completely restricted.
The situation is different for the zero temperature case.
This is obvious from the left graph of Fig.\,\ref{x-a}, where we have plotted the distance between quark and antiquark versus $a$, the rescaled position of the tip of the hanging string from the boundary. 
Even for a very small electric field there is a finite potential barrier at finite temperature, since for a smaller electric field the potential barrier becomes zero at a larger but always finite $x$.
We have also depicted the same thing for the zero temperature case in the right graph of this figure, which shows a behavior similar to the one for confining theories.
This result is consistent with the previous results found for this theory.
It has been shown \cite{linearP} that the D3+D($-1$) theory has a linear rising quark-antiquark potential for zero temperature, a sign for the confinement of quarks.
However, at finite temperature the potential rises linearly but disappears at a finite value of the separation $x$, showing the deconfinement of quarks.

As a consequence of the above argument there must exist a critical electric field below which the Schwinger effect cannot occur.
The existence of such a critical electric field, usually denoted by $E_s$, is intrinsic to the confining phase.
It can be analytically shown that for the theory of our interest at zero temperature $E_s=T_f \sqrt{q}/R^2$, i.e., the potential barrier becomes flat at $x \to \infty$ when $E=T_f \sqrt{q}/R^2$.
It can be easily observed by writing down the total potential at $b=0$ in the following way
\begin{align}
V_{\mathrm{tot}}(x)=2 T_f r_0 a\int^{1/a}_1 \frac{y^2 dy }{\sqrt{y^4-1}}+\left(T_f \frac{q}{R^2\sqrt{(r_0 a)^4+q}}-E\right)x.
\end{align}
When $x \to \infty$ or equivalently $a \to 0$ \footnote{From Eq.\,(\ref{lenthx}) at zero temperature ($b=0$) we simply have $\lim_{a \to 0}x=\lim_{a \to 0} \frac{2 R^2}{r_0 a} \sqrt{1+\frac{q}{(r_0 a)^4}} \int^{1/a}_1 \frac{dy} {y^2\sqrt{y^4-1}}=\infty$.} (see the right graph of Fig.\,\ref{x-a}) the first term in the above equation approaches $2 T_f r_0$ which is constant, i.e., its derivative goes to zero at $x \to \infty$ and the derivative of the second term becomes $T_f \frac{\sqrt{q}}{R^2}-E$, confirming that the electric field should exceed the confining string tension ($T_f \frac{\sqrt{q}}{R^2}$) in order for the pairs to be produced.

An important issue should be noticed here.
For every confining theory there is a mass gap which corresponds to a lower bound on the minimum value of the radial direction $r$ in the gravity side.
A simple analysis shows that the integrand in Eq.\,(\ref{lenthx}) at $b=0$ becomes divergent only at $r=0$, indicating that the lower bound of our theory is at $r=0$, meaning that there is no compactified direction and therefore no scale at zero temperature.
Hence, no geometric transition can be seen here.
This behavior can also be detected using the notion of the entanglement entropy, described below.
%==================================================
\subsection{Entanglement entropy}
In order to calculate the entanglement entropy for the D3+D(-1) background, we follow the method proposed by Klebanov, Kutasov, and Murugan \cite{klebanov}, which is a generalization of the Ryu-Takayanagi conjecture \cite{takayanagi} to nonconformal geometries.
According to their conjecture, the quantum entanglement entropy between a striplike region of length $x$ and its complementary in a $(d+1)$-dimensional quantum field theory corresponds to the classical minimal area of a $d$-dimensional surface $\gamma$ in $AdS_{d+2}$ such that its boundary coincides with the boundary of the striplike region.
For nonconformal geometries in ten dimensions this area can be written as
\begin{align}\label{entroact}
S=\frac{1}{4 G_N^{(10)}}\int d^8 \sigma e^{-2 \Phi}\sqrt{g_{\mathrm{ind}}^{(8)}},
\end{align}
where $G_N^{(10)}$ is the ten-dimensional Newton constant, $g_{\mathrm{ind}}^{(8)}$ is the determinant of the induced metric on the surface $\gamma$, and $\Phi$ is the dilaton field.
By minimizing this action, the entanglement entropy can be obtained.
Using a similar notation as \cite{klebanov}, for a background metric in the form
\begin{align}
ds_{10}^2=\delta(r)\left[\beta(r)dr^2+dx^{\mu}dx_{\mu}\right]+
g_{ij}d\theta^{i}d\theta^{j},
\end{align}
the action (\ref{entroact}) reads
\begin{align}\label{entroact2}
S=\frac{V_2}{4 G_N^{(10)}}\int_{-x/2}^{x/2} d\sigma \sqrt{H(r)}\sqrt{1+\beta(r)(\partial_{\sigma}r)^2},
\end{align}
where $H(r)=e^{-4 \Phi} V_{\mathrm{int}}^2 \delta(r)^3$.
$V_2$ is the volume of the coordinates transverse to the $x$ direction and $V_{\mathrm{int}}$ is the volume of the internal manifold.

In the following we do all the calculations for our theory at zero temperature ($b=0$).
Substituting the functions $H(r)$, $\delta(r)$, and $\beta(r)$ for the geometry of our interest, and calculating the Hamiltonian which is a constant of motion, we find the length of the subsystem, $x$, in terms of the parameters defined in Eq.\;(\ref{parameters}) as follows:
\begin{align}
x(a)=\frac{2 R^2}{r_0 a}\int_{1}^{1/a} \frac{dy} {y^2\sqrt{y^6-1}},
\end{align}
where $a$ denotes the radial position of the tip of the minimized surface in the gravity side that ends on the boundary of the strip with width $x$. 
Using this relation and the action (\ref{entroact2}) we can also find the entanglement entropy in the following form:
\begin{align}\label{con}
S_{C}=\frac{V_2 V_{\Omega_5}}{G_N^{(10)}}R^4(r_0 a)^2\int_{1}^{1/a}dy \frac{y^4} {\sqrt{y^6-1}},
\end{align}
where $V_{\Omega_5}=\frac{V_{\mathrm{int}}}{R^5 e^{5\Phi/2}}$ and the subscript $C$ refers to ``connected.''
Notice that in general another solution is possible, which corresponds to two disconnected surfaces ending at $-x/2$ and $x/2$ in the bulk.
For our metric this solution is found as
\begin{align}\label{discon}
S_{D}=\frac{V_2 V_{\Omega_5}}{G_N^{(10)}}R^4(r_0 a)^2\int_{0}^{1/a}dy \ y.
\end{align}
We subtract this solution, which is constant for any value of the width $x$, from the solution in Eq.\,(\ref{con}) and denote this difference divided by the constant factor $\frac{V_2 V_{\Omega_5}}{G_N^{(10)}}R^4$ by $\Delta S(x)$ in what follows.
\begin{figure}[h]
\begin{center}
\includegraphics[width=6cm]{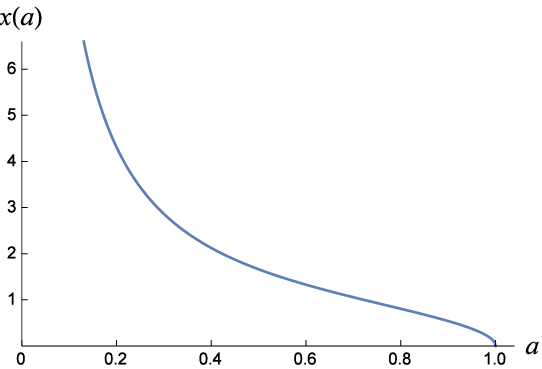}\hspace{.5cm}
\includegraphics[width=6cm]{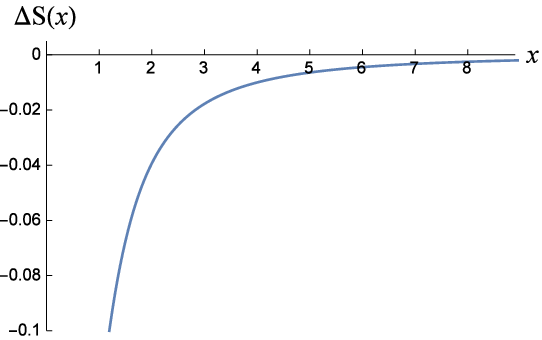}
\end{center}
\caption{\footnotesize 
Left graph: The length of the striplike region versus the locus of the tip of the corresponding surface in the gravity side.
Right graph: The rescaled entanglement entropy versus the size of the entangling region.}
\label{ee}
\end{figure} 

The rescaled entanglement entropy for the field theory of our interest has been drawn numerically in the right panel of Fig.\,\ref{ee}.
In the left panel of this figure we show the length of the entangling region as a function of the locus of the tip of the hypersurface in the bulk, which  ends on the boundary of the striplike region.
Through these figures, it is evident that no confinement/deconfinement phase transition happens in the D3+D($-1$)-brane configuration.
The behavior of the entanglement entropy for our system is exactly like the one for the deconfined field theories (for example see \cite{Sch6,balasubramanian2}).
This result is also obvious from the relations (\ref{con}) and (\ref{discon}) where the dilaton field $\Phi$ and consequently the instanton density $q$ have been removed completely.
Notice that although our theory has no geometric transition and behaves like a deconfined theory as viewed by the entanglement theory, it introduces a critical electric field $E_s$ below which no pairs are produced, which is a property of confined theories.
%*********************************************************************
\section{Pair production rate}
An alternative way to explore the response of a system to an external electric field is to calculate the pair production rate.
This quantity is equivalent to the expectation value of a circular Wilson loop in the $t\textrm{-}x^i$ plane, where $x^i$ refers to each of the spatial coordinates in the field theory. 
This quantity can be evaluated holographically using the calculation of the extremal surface in the bulk which shares the same boundary with the circular Wilson loop.
We devote this section to the calculation of this quantity only for the zero temperature case where we expect the pair production not to occur below the critical electric field $E_s$ found in the previous section.
As stated before when the temperature is nonzero, the pair production rate vanishes at zero electric field and therefore in this case there is no interesting information in this quantity. 
Furthermore, the calculation of the pair production rate at finite temperature is more complicated.
In this case we are dealing with a partial differential equation due to lack of circular symmetry in the $t\textrm{-}x^i$ plane \cite{Sch3}. 

In the following we present the calculation of the production rate briefly. 
More details of calculation can be found in \cite{confin2}.
We work in the Euclidean signature and choose the following ansatz for the bulk coordinates:
\begin{align}
t=\rho(\sigma) \cos \theta, ~~x^1=\rho(\sigma) \sin \theta,~~ z=z(\sigma),
\end{align}
and all other coordinates are constant.
Here $z=\frac{1}{r}$ which is zero at the boundary.
$(\theta, \sigma)$ are the string world-sheet coordinates.
Using this ansatz and choosing $r(\sigma)=\sigma$, the string action reads
\begin{align}\label{ng}
S_{\mathrm{NG}}=&\ 2 \pi R^2 T_f \int_0^x d\rho \frac{\rho \ e^{\Phi(z)/2}}{z^2}\sqrt{1+z'^2},\\
S_{B_2}=& -2 \pi T_f B_{01}\int_0^x d\rho \rho,
\end{align}
where $e^{\Phi(z)}=1+q z^4$ and $x$ denotes the radius of the circular Wilson loop on the D3-brane positioned at $z_0=\frac{1}{r_0}$.
$B_2=B_{01} dt \wedge dx^1$ is an electric 2-form coupled to the NG action.
Then, the pair production rate can be evaluated as $e^{-S}=e^{-S_{\mathrm{NG}}-S_{B_2}}$, where $S_{\mathrm{NG}}$ is the minimized NG action.
The following equation obtained by variation of the action (\ref{ng}) should be satisfied by the function $z(\rho)$
\begin{align}
2 \rho \left(1+z'^2\right)+z\left(1+q z^4\right)\left(z'+z'^3+\rho z''\right)=0.
\end{align}
We solve this equation numerically with the boundary conditions $z(0)=z_c$ and $z'(0)=0$, and the constraint $z'(x)=-\sqrt{\frac{e^{\Phi(z_0)}}{\alpha^2}-1}$.
$z_c$ is the value of $z$ at the tip of the cuplike surface in the bulk.

\begin{figure}[h]
\begin{center}
\includegraphics[width=6cm]{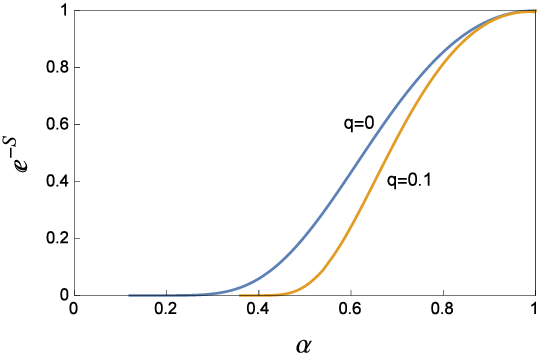}\hspace{.5cm}
\includegraphics[width=6cm]{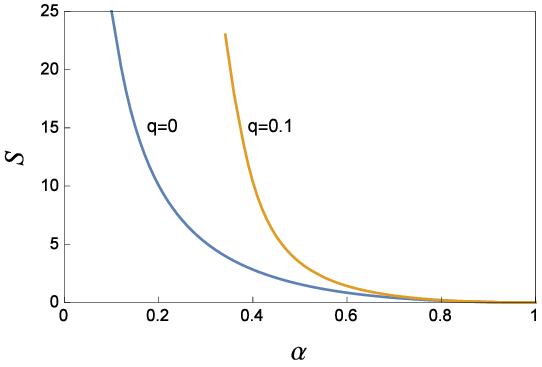}
\end{center}
\caption{\footnotesize 
Left and right graphs respectively show the exponential factor $e^{-S}$, and the classical action $S$ versus $\alpha$.}
\label{rate}
\end{figure}

The exponential factor and the classical action obtained from the above calculations are shown in Fig.\,\ref{rate}.
As can be seen, the exponential factor approaches zero or equivalently the classical action diverges at a certain value of the electric field around $E_s$ obtained by the potential analysis.
For zero instanton density this electric field is zero, i.e., the Schwinger effect occurs for any nonzero, although small, value of the electric field, which is the characteristic of a deconfined system.
This result is in good agreement with the results of the previous section and confirms our numerical calculation.

%************************************************************************
\section{Summary and conclusion}
In this paper we have investigated the Schwinger effect in the presence of the gluon condensation. 
To that purpose we employ the D3+D($-1$)-brane configuration which is a quasiconfining theory with constant gluon condensation.
By quasiconfining we mean that it is deconfined in the viewpoint of gluons but its quark-antiquark potential behaves Coulomb-like for short separations and linearly for medium separations and finally becomes zero for large enough separations depending on the temperature and the gluon condensation.

Using the DBI action for this theory, we have found the critical electric field $E_c$ above which the vacuum decays catastrophically.
For our theory $E_c$ depends on both the gluon condensation and the temperature. 
Then, we have calculated the total potential for a pair of quark-antiquark in the presence of an electric field and find the critical electric field in agreement with the DBI result, as expected.
By comparing the plots of this potential for different values of the D-instanton density we observe that the presence of instantons tends to suppress the pair production and increases the $E_c$ both at finite and zero temperature.
Also, increasing the temperature of this theory decreases the $E_c$ for a fixed gluon condensation.
Interestingly in the case of finite temperature we have not found any critical electric field below which the Schwinger effect cannot occur at all.
That is, at finite temperature pairs are created even in the presence of a very small electric field.
Such a critical electric field is a common feature of all confining field theories and this shows that D3+D($-1$)-brane configuration at finite temperature is deconfined as viewed by an electric field.
This result is expected from the previous investigations such as the behavior of the quark-antiquark potential \cite{linearP}. 
However, we have found such a critical electric field when the temperature is zero, which is consistent with the previous results for the quark-antiquark potential showing the confinement of this theory at zero temperature.
As a final check we have calculated the pair production rate using holography and found consistent results with the ones obtained by the potential analysis.
%*********************************************************************
\section*{Acknowledgement}
L. S. would like to thank the School of Physics of Institute for research in fundamental sciences (IPM) for the research facilities and environment.
The work of M. D. has been supported financially by Research Institute for Astronomy and Astrophysics of Maragha (RIAAM) under research project No. 1/5440-55.
%\appendix
%\section*{Appendix A} \label{Calculation}
%\setcounter{equation}{0}
%\renewcommand{\theequation}{\Alph{section}.\arabic{equation}}


\begin{thebibliography}{9}
\bibitem{Schwinger}
  J. S. Schwinger,
  ``On gauge invariance and vacuum polarization,''
  Phys. Rev. {\bf 82}, 664 (1951).

\bibitem{manton1}
I. K. Affleck, O. Alvarez and N. S. Manton, ``Pair production at strong coupling in weak external fields,'' Nucl. Phys. B {\bf 197}, 509 (1982).

%\bibitem{manton2}
%I. K. Affleck and N. S. Manton, ``Monopole pair production in a magnetic field,'' Nucl. Phys. B {\bf 194}, 38 (1982).

\bibitem{string1}
E. S. Fradkin and A. A. Tseytlin, ``Quantum string theory effective action,'' Nucl. Phys. B {\bf 261}, 1 (1985).

\bibitem{string2}
C. Bachas and M. Porrati, ``Pair creation of open strings in an electric field,'' Phys. Lett. B
{\bf 296}, 77 (1992) \href{https://arxiv.org/abs/hep-th/9209032}{[arXiv:hep-th/9209032]}.

\bibitem{Maldacena1}
  J.~M.~Maldacena,
  ``The large N limit of superconformal field theories and gravity,''
  Adv.\ Theor.\ Math.\ Phys.\  {\bf 2}, 231 (1998)
  [Int.\ J.\ Theor.\ Phys.\  {\bf 38}, 1113 (1999)]
  \href{http://arxiv.org/abs/hep-th/9711200}{[arXiv:hep-th/9711200]}.

\bibitem{Maldacena2}
  S.~S.~Gubser, I.~R.~Klebanov and A.~M.~Polyakov,
  ``Gauge theory correlators from non-critical string theory,''
  Phys.\ Lett.\  B {\bf 428}, 105 (1998)
  \href{http://arxiv.org/abs/hep-th/9802109}{[arXiv:hep-th/9802109]}.

\bibitem{Maldacena3}
  E.~Witten,  ``Anti-de Sitter space and holography,''
  Adv.\ Theor.\ Math.\ Phys.\  {\bf 2}, 253 (1998)
  \href{http://arxiv.org/abs/hep-th/9802150}{[arXiv:hep-th/9802150]}.

\bibitem{Solana} 
  J.~Casalderrey-Solana, H.~Liu, D.~Mateos, K.~Rajagopal and U.~A.~Wiedemann,
  ``Gauge/string duality, hot QCD and heavy ion collisions,''
    \href{http://arxiv.org/abs/1101.0618}{[arXiv:1101.0618 [hep-th]]}.

\bibitem{semenoff}
G. W. Semenoff and K. Zarembo, ``Holographic Schwinger effect,''
Phys. Rev. Lett. {\bf 107}, 171601 (2011) \href{https://arxiv.org/abs/1109.2920}{[arXiv:1109.2920 [hep-th]]}.

\bibitem{potential}
Y. Sato and K. Yoshida, ``Potential analysis in holographic Schwinger effect,'' JHEP {\bf 08}, 002 (2013) \href{https://arxiv.org/abs/1304.7917}{[arXiv:1304.7917 [hep-th]]}.

\bibitem{Sch1}
Y. Sato and K. Yoshida, ``Universal aspects of holographic Schwinger effect in general backgrounds,'' JHEP {\bf 12}, 051 (2013) \href{https://arxiv.org/abs/1309.4629}{[arXiv:1309.4629 [hep-th]]}.

\bibitem{Sch2}
Y. Sato and K. Yoshida, ``Holographic description of the Schwinger effect in electric
and magnetic field,'' JHEP {\bf 04}, 111 (2013) \href{https://arxiv.org/abs/1303.0112}{[arXiv:1303.0112 [hep-th]]}.

\bibitem{Sch3}
S. Bolognesi, F. Kiefer and E. Rabinovici, ``Comments on critical electric and magnetic fields from holography,'' JHEP {\bf 01}, 174 (2013) \href{https://arxiv.org/abs/1210.4170}{[arXiv:1210.4170 [hep-th]]}.

\bibitem{confin1}
Y. Sato and K. Yoshida, ``Holographic Schwinger effect in confining phase,'' JHEP {\bf 09}, 134 (2013) \href{https://arxiv.org/abs/1306.5512}{[arXiv:1306.5512 [hep-th]]}.

\bibitem{confin2}
D. Kawai, Y. Sato and K. Yoshida, ``The Schwinger pair production rate in confining
theories via holography,'' Phys. Rev. D {\bf 89}, 101901 (2014) \href{https://arxiv.org/abs/1312.4341}{[arXiv:1312.4341 [hep-th]]}.

\bibitem{Sch4}
K. Hashimoto, S. Kinoshita, K. Murata and T. Oka, ``Electric field quench in AdS/CFT,'' JHEP {\bf 09}, 126 (2014) \href{https://arxiv.org/abs/1407.0798}{[arXiv:1407.0798 [hep-th]]}.

\bibitem{Sch5}
W. Fischler, P. H. Nguyen, J. F. Pedraza and W. Tangarife, ``Holographic Schwinger
effect in de Sitter space,'' Phys. Rev. D {\bf 91}, 086015 (2015) \href{https://arxiv.org/abs/1411.1787}{[arXiv:1411.1787 [hep-th]]}.

\bibitem{Sch6}
M. Ghodrati, ``Schwinger effect and entanglement entropy in confining geometries,'' Phys. Rev. D {\bf 92}, 065015 (2015) \href{https://arxiv.org/abs/1506.08557}{[ 	arXiv:1506.08557 [hep-th]]}.

\bibitem{Sch7}
S-J. Zhang and E. Abdalla, ``Holographic Schwinger effect in a confining background with Gauss-Bonnet corrections,'' Gen. Rel. Grav. {\bf 48}, 60 (2016) \href{https://arxiv.org/abs/1508.03364}{[arXiv:1508.03364 [hep-th]]}.

\bibitem{confinrev}
D. Kawai, Y. Sato and K. Yoshida, ``A holographic description of the Schwinger effect in a confining gauge theory,'' Int. J. Mod. Phys. A {\bf 30}, 1530026 (2015) \href{https://arxiv.org/abs/1504.00459}{[arXiv:1504.00459 [hep-th]]}.

\bibitem{dual-ins1}
 T. Banks and M. B. Green, ``Non-perturbative effects in $\mathrm{AdS}_5 \times \mathrm{S}^5$ string theory and $d=4$ SUSY Yang-Mills,'' JHEP {\bf 05}, 002 (1998) \href{https://arxiv.org/abs/hep-th/9804170}{[hep-th/9804170]}.

\bibitem{dual-ins2}
V. Balasubramanian, P. Kraus, A. Lawrence and S. P. Trivedi, ``Holographic probes of anti-de Sitter spacetimes,'' Phys. Rev. D {\bf 59}, 104021 (1999) \href{https://arxiv.org/abs/hep-th/9808017}{[arXiv:hep-th/9808017]}.

\bibitem{dual-ins3}
O. Aharony, S. S. Gubser, J. Maldacena, H. Ooguri and Y. Oz, ``Large $N$ field theories, string theory and gravity,'' Phys. Rept. {\bf 323}, 183 (2000) \href{https://arxiv.org/abs/hep-th/9905111}{[arXiv:hep-th/9905111]}.

\bibitem{instanton1}
H. Liu and A. A. Tseytlin, ``D3-brane-D-instanton configuration and $N$ = 4 super YM theory in constant self-dual background,'' Nucl. Phys. {\bf 553}, 231 (1999) \href{https://arxiv.org/abs/hep-th/9903091}{[arXiv:hep-th/9903091]}.

\bibitem{instanton2}
B. Gwak, M. Kim, B.-H. Lee, Y. Seo and S.-J. Sin, ``Holographic D instanton liquid and chiral transition,'' Phys. Rev. D {\bf 86}, 026010 (2012) \href{https://arxiv.org/abs/1203.4883}{[arXiv:1203.4883 [hep-th]]}.

\bibitem{linearP}
K. Ghoroku, T. Sakaguchi, N. Uekusa and M. Yahiro, ``Flavor quark at high temperature
from a holographic model,'' Phys. Rev. D {\bf 71}, 106002 (2005) \href{https://arxiv.org/abs/hep-th/0502088}{[arXiv:hep-th/0502088]}.

\bibitem{klebanov}
I. R. Klebanov, D. Kutasov and A. Murugan, ``Entanglement
as a probe of confinement,'' Nucl. Phys. B {\bf 796}, 274 (2008) \href{https://arxiv.org/abs/0709.2140}{[arXiv:0709.2140 [hep-th]]}.

\bibitem{takayanagi}
S. Ryu and T. Takayanagi, ``Holographic derivation of entanglement entropy from AdS/CFT,'' Phys. Rev. Lett. {\bf 96}, 181602 (2006)
\href{https://arxiv.org/abs/hep-th/0603001}{[arXiv:hep-th/0603001]}.

\bibitem{balasubramanian2}
V. Balasubramanian,  A. Bernamonti, J. de Boer, N. B. Copland, B. Craps, E. Keski-Vakkuri, B. Müller, A. Schäfer, M. Shigemori and W. Staessens,
``Holographic thermalization,''
Phys. Rev. D {\bf 84}, 026010 (2011)
 \href{http://arxiv.org/abs/1103.2683}{[arXiv:1103.2683 [hep-th]]}.
\end{thebibliography}
 \end{document}